# A Fast and Accurate Semi-Empirical Approach for Hydrogen-Exchange Kinetic Isotope Effect Evaluation

*Mikhail Rudenko[1], Artem Eliseev[1,2], Artem Mitrofanov[1,2], Stepan Kalmykov[1]*

[1]Department of Chemistry, Lomonosov Moscow State University, Moscow 119991, Russia

[2]MSU Institute for Artificial Intelligence, Lomonosov Moscow State University, Moscow 119192, Russia

**ABSTRACT**

The kinetic isotope effect (KIE) is essential in various chemical applications from reaction mechanism studies to tritium removal from water. Traditional KIE evaluation relies on experimental measurements or computational approaches like density functional theory (DFT), which are often costly and inaccurate. Here, we present a novel semi-empirical method for rapid and precise KIE estimation in proton-exchange reactions. By refining transition state identification through an iterative surface scan, our approach significantly improves accuracy while maintaining computational efficiency. Benchmarking against experimental data demonstrates superior performance compared to both DFT and conventional semi-empirical methods. Additionally, validation with tritium exchange reactions confirms its robustness. The computational implementation is freely available, facilitating its integration into future research.

**TOC GRAPHICS**

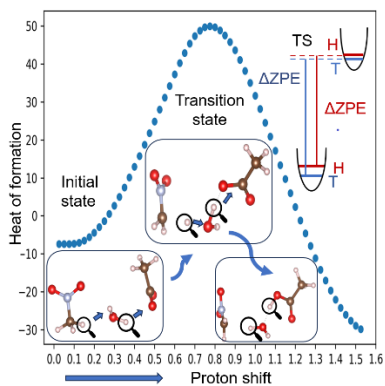

The kinetic isotope effect (KIE) – defined as the ratio of reaction rate constants for different isotopes - is a fundamental physicochemical phenomenon employed across diverse fields from reaction mechanism studies[1–5] to biochemistry[6–8], catalysis[9,10] and electrochemistry[11,12]. The use of stable, isotopically substituted materials as tracers in biological systems has also seen rapid growth, driven by recent technological advances[13]. One of the most critical potential application of KIE is the water purification, particularly the removal of various isotopes, especially tritium, which typically involves energy-intensive and time-consuming methods[14,15]. To address these challenges, alternative methods of water purification are being developed, including the use of bioreactors [16], MOFs [17] or membranes [18]. The rational usage of KIE that may be the key to achieve the greatest cost reduction in this process[19].

KIE arises from differences in nuclear structures among isotopes, leading to variations in bond formation or dissociation energies. Materials exhibiting KIE in proton transfer rates between isotopes (e.g., $^1H$ and $^3H$) in water purification processes offer promising, energy-efficient alternatives. However, the experimental determination of KIE values remains resource-intensive, time-consuming, and potentially hazardous in a case of radioactive isotopes. A viable alternative is to employ theoretical approaches based on quantum chemistry methods, such as density functional theory (DFT)[20]. However, these methods are computationally demanding and often lack accuracy. Here, we introduce a semi-empirical approach that facilitates a faster and more accurate KIE estimation for proton-exchange reactions.

**Computational Details**

The theoretical KIE was estimated following a specific computational scheme (see SI 1). Initial molecular geometries were optimized through molecular mechanics (MM) in GABEDIT [21], utilizing Amber parameters. Subsequent optimization used the same level of theory as the subsequent calculations (see below). The KIE calculations employed only the zero-point energy (ZPE) contribution, (see SI 3). KIE values were derived using the following equation:

$$KIE = \frac{k(H)}{k(D)} = \exp\left(\frac{-([(ZPE_{TS(H)} - ZPE_{R(H)})] - [(ZPE_{TS(D)} - ZPE_{R(D)})])}{RT}\right),$$

where k was the reaction rate constant, TS denoted the transition state, and R the reactant state. The corresponding isotope was indicated in brackets. Thermodynamic calculations were performed at 298 K, with validation using a comprehensive database of experimental H/D KIE values for various reactions (see the whole list of reactions in the Supplementary Information, Section 1). Deuterium was selected as a model isotope due to the availability of extensive experimental data, though results remain applicable to tritium due to similar KIE mechanisms.

Quantum-chemical calculations were conducted using ORCA 5.0.4 [22] and OpenMOPAC v22.1.1 [23] on two INTEL GOLD 6338 processors in 100-thread configurations.

**DFT-Based Approach**

Since DFT methods are widely used in quantum chemical calculations, we began by evaluating their performance. The B3LYP [24] functional with Grimme's dispersion correction D3 [25], favored for organic molecules due to high accuracy with relatively low complexity, was employed for geometry optimization and ZPE calculation. The def2-TZVP basis set was applied [26]. Calculations were optimized for proton exchange reactions occurring through one or more water molecules, with the COSMO solvent model employed [27]. The computational procedure followed this scheme:

1. The initial geometry of the system of each reaction was optimized without coordinates constraints. Then ZPE(R) value was calculated by internal OPT ORCA [22] module.

2. The proton was moved to the acceptor molecule and the reaction product geometry was obtained. The geometry was also optimized and its ZPE was calculated.

3. The transition state was determined using the Nudged Elastic Band (NEB) method with 6 images configuration, with subsequent geometry optimization and ZPE(TS) calculation.

4. ZPE for deuterium-substituted system was calculated by replacing the corresponding hydrogen atom in the optimized system of reactants (R) and the transition state (TS). Both geometries were not optimized after substitution since the geometry of its ground state does not depend on the isotope mass in the computational approximation.

**Classic Semi-Empirical Approach**

Reactant and product geometries were optimized in OpenMOPAC[23] (PM7 Hamiltonian [28]), and the transition state was located using the SADDLE method. PM7 is capable of showing accuracy at the level of DFT, while the duration of calculations is strikingly shorter [29]. Although PM7 was proposed in 2013, it is still relevant, since many modern models, including those using machine learning [30] and protein models [31], based on it show excellent results [32]. The whole scheme was:

1. The initial geometry of each of the reactants was separately optimized, then the molecules were positioned with the reaction groups pointing to each other direction and the geometry of the system was optimized again. No atoms were fixed during this step. The ZPE(R) value was determined using THERMO module of OpenMOPAC after optimization.

2. The exchanged proton was transferred to the acceptor molecule and the products geometry was also optimized using the PM7 Hamiltonian and system ZPE was calculated.

3. The transition state was obtained using the PM7 SADDLE method and system ZPE(TS) was calculated.

4. ZPE for deuterium-substituted system was calculated by replacing the corresponding hydrogen atom in the optimized system of reactants (R) and transition state (TS), without additional geometry optimization.

**Proposed Semi-Empirical Approach**

The approach we propose improves the search for the transition state by iteratively shifting the hydrogen atom. The hydrogen atom was shifted in the direction of the shift vector connecting this atom and the acceptor atom of the base (e.g. O or N). To implement this approach, we could have used the preservation of the total distance between the Donor-Hydrogen-Acceptor atoms, but we chose, in our opinion, a more transparent and variable approach consisting in fixing certain coordinates of the atoms. The OpenMOPAC program allows fixing one of the coordinates during geometry optimization, in our case X-axis, allowing the atom to shift in the other two planes. To implement it, at the beginning of the calculation, the coordinate system was rotated using multiplication by the rotation and translation matrix, as a result of which the hydrogen atom ends up at the point (0, 0, 0), and the shift vector coincides with the X-axis. The approach is generally based on the idea of scanning a relaxed surface.

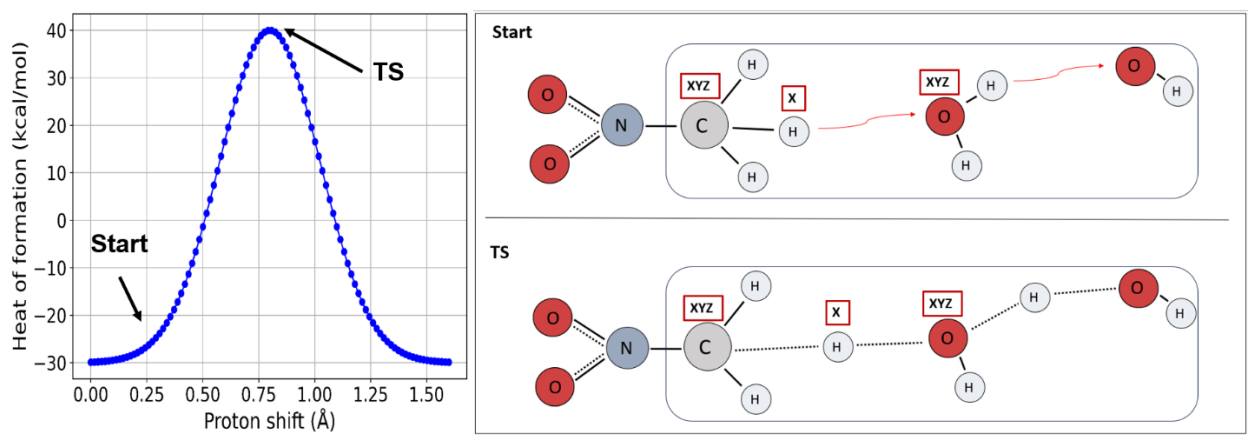

**Figure 1.** Reaction profile for the exchange of protons between CH3NO2 and OH, with the geometries of the system in the initial (start) and transition states (TS) shown on the right.

1. The hydrogen atom was shifted by 0.025 Å step along the X-axis (further decrease of the step did not lead to a noticeable improvement in the results, while the calculation time increased respectively). The geometry was optimized, with a fixed position of hydrogen on the X-axis after each step.

2. The Hessian and ZPE were calculated and stored together with the optimized geometry and heat of formation (HOF) value.

3. The protium isotope was replaced with the target one. The zero-point energy of the isotope-substituted system was calculated using previously saved Hessian.

4. Steps 1-3 were repeated until the distance between hydrogen atom and acceptor (R) became less than 1 Å. Further displacement of hydrogen led only to an increase in energy due to interatomic repulsion.

5. The transition state was determined from the obtained reaction energy profile (Figure 2), as a point with the highest energy. The KIE was calculated using equation 1 from previously calculated ZPE of the initial and transition states for all isotopes.

This algorithm was implemented in Python3 (OpenMOPAC package required), with the code available at https://github.com/SmartChemDesign/kie. Users need only specify the initial geometry, the exchangeable hydrogen atom, and the acceptor atom. The repository also includes all calculation results, geometries, and ZPE values reported in the current article.

We compared the efficiency of the methods for a set of 20 reactions with experimentally known KI values. All reactions used for verification are given in SI 5.

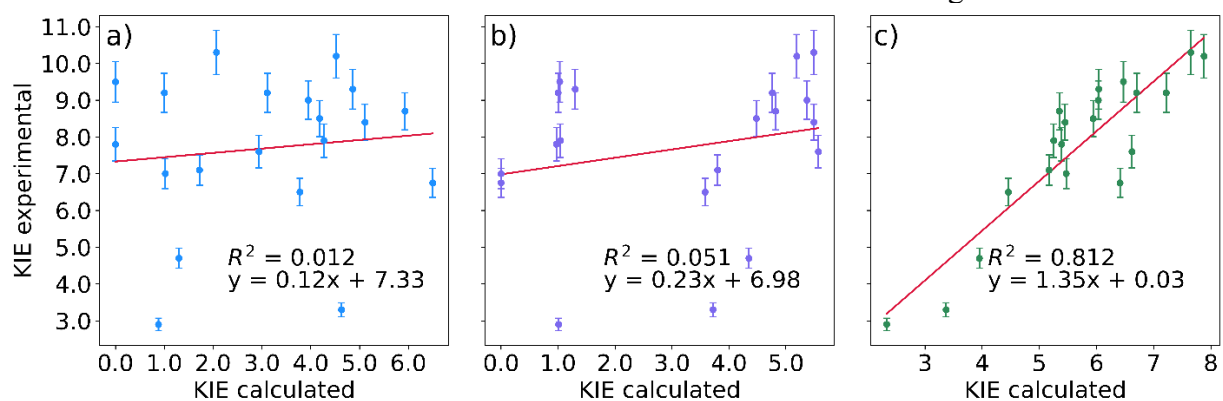

**Figure 2.** Correlation between the experimental data and calculated KIE value using DFT (a), SADDLE MOPAC implementation (b) and our proposed algorithm (c).

Our DFT calculations confirmed the inaccuracy and lack of correlation between computed and experimental values (Figure 2a), arising from high variability in transition state determination, particularly in reactions involving water molecules, compromised DFT accuracy, taking into account average calculation time of 11.3 hours, the results can be at least described as unsatisfactory. Semi-empirical methods, though faster, yielded considerably lower accuracy than DFT (Figure 2b). The reason lies in the fact, that semi-empirical methods are limited by their criteria for transition state identification (relying on fewer vibrational mode checks). In reactions with water as a bridge molecule this limitation frequently leads to unrealistic geometries.

On the other hand, the proposed approach demonstrated significantly better accuracy with highly reduced calculation time, compared to DFT. Although the KIE values were usually underestimated due to our focus on the primary KIE, due to the fact that we do not take into account the secondary KIE, which slightly increases the overall KIE value, namely the total KIE value is determined experimentally. However, the use of linear approximation allowed us to obtain more accurate predictive results (Figure 2c).

The average mean absolute error and calculation time, normalized by electron number for each method are summarized in Table 1. The proposed approach exhibits the lowest error value and the computation time comparable to the original semiempirical approach.

**Table 1.** Comparison of the average calculation time, in minutes per number of electrons in system, and the average absolute error of all methods.

| Method | Average MAE | Average Time, min per electron |
|---|---|---|
| DFT | 4.33 | 4.730 |
| Semi-empirical | 4.97 | 0.031 |
| Proposed | 2.13 | 0.054 |

To demonstrate the extensibility of the proposed approach, we collected data on tritium exchange reactions and compared the theoretical predictions with experimental results. The corresponding values are presented in Table 2. Notably, in most cases, the theoretical predictions fell within the experimental uncertainty limits. The correlation between the theoretical and experimental data is illustrated in Figure 3.

**Table 2.** The tritium/protium KIE.

| Molecule | Exp. | Calc. |
|---|---|---|
| acetone[33] | 63.89±4.3 | 53.55 |
| acetophenon[34] | 15.10±2.0 | 16.60 |
| 2-phenylacetophenone[35] | 10.20±0.32 | 10.39 |
| o-bromo_acetophenon[36] | 11.90±3.0 | 11.52 |
| o-fluoro_acetophenon[36] | 13.00±3.0 | 12.51 |
| o-iodo_acetophenon[36] | 12.50±3.0 | 9.84 |
| o-methoxy_acetophenon[36] | 19.80±5.0 | 13.14 |
| o-methyl_acetophenon[36] | 16.30±4.0 | 14.54 |

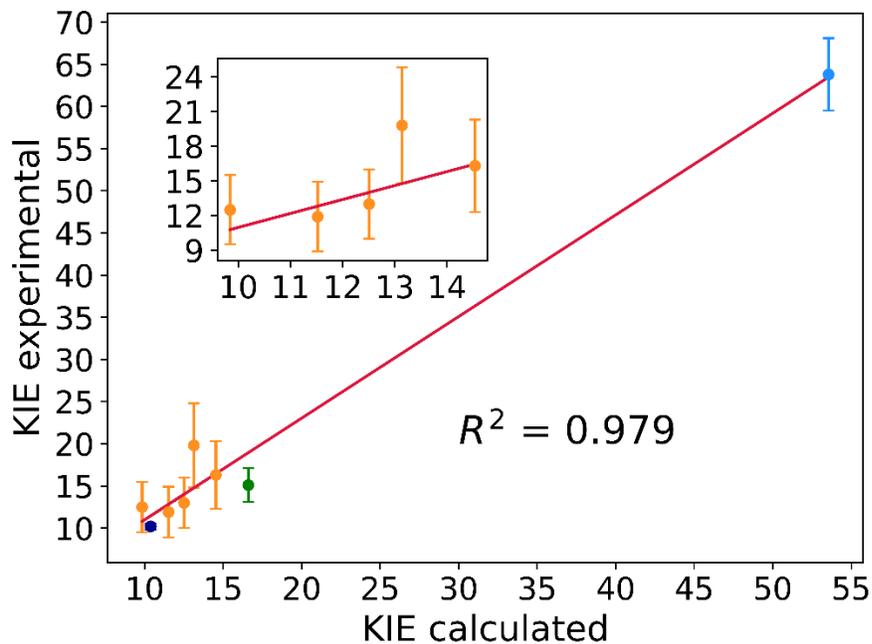

**Figure 3.** Correlation between experimental data and calculated KIE values using proposed algorithm. Data from different articles are marked in different colors. Their linear approximation is shown as a red line.

As shown in Figure 3, our method exhibits excellent correlation with the experimental data, demonstrating its reliability in predicting kinetic isotope effects (KIE). Notably, the data point corresponding to the largest KIE values appears to have the greatest statistical influence on the overall correlation. When analyzing a subset of the data obtained under the same conditions separately, the correlation coefficient decreases; however, the approximating trend remains within the experimental confidence interval. This consistency across different data sets underscores the robustness of the proposed approach.

Furthermore, the strong agreement between theoretical predictions and experimental results highlights the accuracy and predictive power of our method. Unlike conventional computational approaches, which often suffer from high computational costs and reduced accuracy, our semi-empirical method provides a balance between efficiency and precision. These findings confirm the high applicability of the developed approach for modeling proton-exchange reactions, particularly in scenarios where rapid and reliable KIE estimations are required.

This work introduces a novel semi-empirical approach for KIE estimation in organic proton-exchange reactions, offering superior to traditional DFT calculations accuracy and retaining computational efficiency of common semi-empirical methods. We believe reported computational approach will pave the way in the development of improved tritium sorption and separation techniques and materials.

ASSOCIATED CONTENT

**Data Availability Statement**

The code is available at https://github.com/SmartChemDesign/kie.

**Supporting Information**

The Supporting Information is available in SI_eq.pdf

ACKNOWLEDGMENT

The study was supported by the Russian Science Foundation (grant no. 23-73-30006).

REFERENCES

(1) Westaway, K. C.; Pham, T. Van; Fang, Y.-R. *Using Secondary R Deuterium Kinetic Isotope Effects To Determine the Symmetry of S N 2 Transition States*; 1997, DOI: 10.1021/ja962088f.

(2) Mao, Z.; Campbell, C. T. Kinetic Isotope Effects: Interpretation and Prediction Using Degrees of Rate Control.*ACS Catal.***2020**, *10* (7), 4181–4192, DOI: 10.1021/acscatal.9b05637.

(3) Cleland, W. W. The Use of Isotope Effects to Determine Enzyme Mechanisms.*J. Biol. Chem.***2003**, *278* (52), 51975–51984, DOI: 10.1074/jbc.X300005200.

(4) Gómez-Gallego, M.; Sierra, M. A. Kinetic Isotope Effects in the Study of Organometallic Reaction Mechanisms.*Chem. Rev.***2011**, *111* (8), 4857–4963, DOI: 10.1021/cr100436k.

(5) Simmons, E. M.; Hartwig, J. F. On the Interpretation of Deuterium Kinetic Isotope Effects in C−H Bond Functionalizations by Transition-Metal Complexes.*Angew. Chemie Int. Ed.***2012**, *51* (13), 3066–3072, DOI: 10.1002/anie.201107334.

(6) Ji, L.; Zhang, H.; Ding, W.; Song, R.; Han, Y.; Yu, H.; Paneth, P. Theoretical Kinetic Isotope Effects in Establishing the Precise Biodegradation Mechanisms of Organic Pollutants.*Environ. Sci. Technol.***2023**, *57* (12), 4915–4929, DOI: 10.1021/acs.est.2c04755.

(7) Shao, L.; Hewitt, M. C. The Kinetic Isotope Effect in the Search for Deuterated Drugs.*Drug News Perspect.***2010**, *23* (6), 398, DOI: 10.1358/dnp.2010.23.6.1426638.

(8) Sharma, R.; Strelevitz, T. J.; Gao, H.; Clark, A. J.; Schildknegt, K.; Obach, R. S.; Ripp, S. L.;

Spracklin, D. K.; Tremaine, L. M.; Vaz, A. D. N. Deuterium Isotope Effects on Drug Pharmacokinetics. I. System-Dependent Effects of Specific Deuteration with Aldehyde Oxidase Cleared Drugs. *Drug Metab. Dispos.* **2012**, *40* (3), 625–634, DOI: 10.1124/dmd.111.042770.

(9) Tse, E. C. M.; Varnell, J. A.; Hoang, T. T. H.; Gewirth, A. A. Elucidating Proton Involvement in the Rate-Determining Step for Pt/Pd-Based and Non-Precious-Metal Oxygen Reduction Reaction Catalysts Using the Kinetic Isotope Effect. *J. Phys. Chem. Lett.* **2016**, *7* (18), 3542–3547, DOI: 10.1021/acs.jpclett.6b01235.

(10) Meemken, F.; Rodríguez-García, L. Revealing Catalytically Relevant Surface Species by Kinetic Isotope Effect Spectroscopy: H-Bonding to Ester Carbonyl of Trans-Ethyl Pyruvate Controls Enantioselectivity on a Cinchona-Modified Pt Catalyst. *J. Phys. Chem. Lett.* **2018**, *9* (5), 996–1001, DOI: 10.1021/acs.jpclett.7b03360.

(11) Rebollar, L.; Intikhab, S.; Snyder, J. D.; Tang, M. H. Kinetic Isotope Effects Quantify PH-Sensitive Water Dynamics at the Pt Electrode Interface. *J. Phys. Chem. Lett.* **2020**, *11* (6), 2308–2313, DOI: 10.1021/acs.jpclett.0c00185.

(12) Cai, S.; Bai, T.; Chen, H.; Fang, W.; Xu, Z.; Lai, H.; Huang, T.; Xu, H.; Chu, X.; Ling, J.; Gao, C. Heavy Water Enables High-Voltage Aqueous Electrochemistry via the Deuterium Isotope Effect. *J. Phys. Chem. Lett.* **2020**, *11* (1), 303–310, DOI: 10.1021/acs.jpclett.9b03267.

(13) Stürup, S.; Hansen, H. R.; Gammelgaard, B. Application of Enriched Stable Isotopes as Tracers in Biological Systems: A Critical Review. *Analytical and Bioanalytical Chemistry*. January 2008, pp 541–554, DOI: 10.1007/s00216-007-1638-8.

(14) Shibuya, S.; Matsushima, H.; Ueda, M. Study of Deuterium Isotope Separation by PEFC. *J. Electrochem. Soc.* **2016**, *163* (7), F704–F707, DOI: 10.1149/2.1321607jes.

(15) Lehto, J.; Koivula, R.; Leinonen, H.; Tusa, E.; Harjula, R. Removal of Radionuclides from Fukushima Daiichi Waste Effluents. *Sep. Purif. Rev.* **2019**, *48* (2), 122–142, DOI: 10.1080/15422119.2018.1549567.

(16) Li, L. Process Tritiated Water through a Bioreactor: An Ideal Model for Dealing the Fukushima Wastewater. *Anal. Chem. A J.* **2023**, *2* (1), 108–114, DOI: 10.23977/analc.2023.020114.

(17) Rethinasabapathy, M.; Ghoreishian, S. M.; Hwang, S. K.; Han, Y. K.; Roh, C.; Huh, Y. S. Recent Progress in Functional Nanomaterials towards the Storage, Separation, and Removal of Tritium. *Adv. Mater.* **2023**, *35* (48), DOI: 10.1002/adma.202301589.

(18) Park, C. W.; Kim, S. W.; Kim, H. J.; Jeong, E.; Yoon, I. H. Separation Behavior of Hydrogen Isotopes via Water Pervaporation Using Proton Conductive Membranes. *Environ. Sci. Water Res. Technol.* **2024**, No. 11, 2787–2795, DOI: 10.1039/d4ew00330f.

(19) Kim, H.; Kumar Singh, B.; Um, W. Tritium Separation from Radioactive Wastewater by

Hydrogen Isotope-Selective Exchange of Hydrogen-Bonded Fluorine. *J. Ind. Eng. Chem.* **2023**, *121*, 264–274, DOI: 10.1016/j.jiec.2023.01.030.

(20) Christensen, N. J.; Fristrup, P. Kinetic Isotope Effects (KIE) and Density Functional Theory (DFT): A Match Made in Heaven? *Synlett* **2015**, *26* (4), 508–513, DOI: 10.1055/s-0034-1380097.

(21) Allouche, A. R. Gabedita - A Graphical User Interface for Computational Chemistry Softwares. *J. Comput. Chem.* **2011**, *32* (1), 174–182, DOI: 10.1002/jcc.21600.

(22) Neese, F.; Wennmohs, F.; Becker, U.; Riplinger, C. The ORCA Quantum Chemistry Program Package. *J. Chem. Phys.* **2020**, *152* (22), DOI: 10.1063/5.0004608.

(23) Stewart, J. J. P., Klamt, A., Thiel, W., Danovich, D., Rocha, G. B., Gieseking, R. L., Moussa, J. E., Kurtz, H. A., Korambath, P., Merz, K. M., & Wang, B. MOPAC. 2022, DOI: 10.5281/zenodo.6728590.

(24) Tirado-Rives, J.; Jorgensen, W. L. Performance of B3LYP Density Functional Methods for a Large Set of Organic Molecules. *J. Chem. Theory Comput.* **2008**, *4* (2), 297–306, DOI: 10.1021/ct700248k.

(25) Grimme, S.; Antony, J.; Ehrlich, S.; Krieg, H. A Consistent and Accurate Ab Initio Parametrization of Density Functional Dispersion Correction (DFT-D) for the 94 Elements H-Pu. *J. Chem. Phys.* **2010**, *132* (15), DOI: 10.1063/1.3382344.

(26) Zheng, J.; Xu, X.; Truhlar, D. G. Minimally Augmented Karlsruhe Basis Sets. *Theor. Chem. Acc.* **2011**, *128* (3), 295–305, DOI: 10.1007/s00214-010-0846-z.

(27) Klamt, A. The COSMO and COSMO-RS Solvation Models. *Wiley Interdiscip. Rev. Comput. Mol. Sci.* **2011**, *1* (5), 699–709, DOI: 10.1002/wcms.56.

(28) Stewart, J. J. P. Optimization of Parameters for Semiempirical Methods VI: More Modifications to the NDDO Approximations and Re-Optimization of Parameters. *J. Mol. Model.* **2013**, *19* (1), 1–32, DOI: 10.1007/s00894-012-1667-x.

(29) Vyboishchikov, S. F.; Voityuk, A. A. Solvation Free Energies for Aqueous and Nonaqueous Solutions Computed Using PM7 Atomic Charges. *J. Chem. Inf. Model.* **2021**, *61* (9), 4544–4553, DOI: 10.1021/acs.jcim.1c00885.

(30) García-Andrade, X.; García Tahoces, P.; Pérez-Ríos, J.; Martínez Núñez, E. Barrier Height Prediction by Machine Learning Correction of Semiempirical Calculations. *J. Phys. Chem. A* **2023**, *127* (10), 2274–2283, DOI: 10.1021/acs.jpca.2c08340.

(31) Stewart, J. J. P.; Stewart, A. C. A Semiempirical Method Optimized for Modeling Proteins. *J. Mol. Model.* **2023**, *29* (9), 1–18, DOI: 10.1007/s00894-023-05695-1.

(32) Sun, Q.; Gieseking, R. L. M. Parametrization of the PM7 Semiempirical Quantum Mechanical Method for Silver Nanoclusters. *J. Phys. Chem. A* **2022**, *126* (37), 6558–6569,

DOI: 10.1021/acs.jpca.2c05782.

(33) Jones, J. R. *Rates of Abstraction of Hydrogen, Deuterium and Tritium from Acetone in Alkaline Media*; 1965.

(34) Jones, J. R.; Marks, R. E.; Subba Rao, S. C. *Kinetic Isotope Effects. Part 2. - Rates of Abstraction of Hydrogen and Tritium from Acetophenone and Some Para- and Meta-Substituted Acetophenones in Alkaline Media*; **1967**; Vol. 63, DOI: 10.1039/TF9676300111.

(35) C. Gardner Swain­Edward C. Stivers­Joseph F. Reuwer Jr.­Lawrence J. Schaad. Use of Hydrogen Isotope Effects to Identify the Attacking Nucleophile in the Enolization of Ketones Catalyzed by Acetic Acid1-3.*J. Am. Chem. Soc.***1958**, *80* (21), 5885–5893.

(36) Jones, B. Y. J. R.; Marks, R. E.; Subba Rao, S. C. *Kinetic Isotope Effects. Part 4. - Bromination and Detritiation of Some Ortho-Substituted Acetophenones in Alkaline Media*; **1967**; Vol. 63, DOI: 10.1039/TF9676300993.